\documentclass[%
 reprint,
nofootinbib,
 amsmath,amssymb,
 aps,
]{revtex4-2}

\usepackage{graphicx}
\usepackage{dcolumn}
\usepackage{bm}
\usepackage{braket}
\usepackage{comment}
\usepackage{siunitx}
\usepackage[colorlinks=true, allcolors=blue]{hyperref}

\begin{document}

\preprint{APS/123-QED}

\title{Stochastic effect on ultralight dark matter searches with astronomical polarized sources}

\author{Kyohei Yamada}
\email{kyoheiy@princeton.edu}
\affiliation{Joseph Henry Laboratories of Physics, Jadwin Hall, Princeton University, Princeton, NJ 08544, USA\looseness=-1}

\date{\today}

\begin{abstract}
Ultralight axion-like particles (ALP) are the well-motivated candidates for the dark matter. 
In the standard halo model, the ultralight galactic dark matter behaves as a stochastic Gaussian random field, and the polarization angle of the linearly polarized light propagating through the field stochastically oscillates proportional to the field difference between the observer and the source. The statistics of this signal depend on the source size and its distance from the observer relative to the ALP Compton length and coherence length, and therefore on the ALP mass.
We present a unified statistical framework for the polarization-oscillation searches with light sources of arbitrary size and distance.
We classify the statistics of the stochastic oscillation amplitude into four distinct regimes and provide a seamless description including the transitional regimes between them.
Our framework reproduces the signal models in previously known cases including the laboratory-scale searches and searches using pulsars and the cosmic microwave background. We provide a likelihood prescription that incorporates the stochasticity of the signal, and show that uncertainties in source distance must be carefully accounted for, as they can alter the resulting constraints on the ALP-photon coupling by up to a factor of three.
Our results provide a consistent basis for interpreting and comparing a broad range of ALP searches with astronomical polarized sources.
\end{abstract}

\maketitle
\section{Introduction}
Ultralight axion-like particles (ALP) with masses as low as $m_a \sim 10^{-22}$~eV are well-motivated dark matter (DM) candidates \cite{WITTEN1984351, PhysRevD.81.123530, FuzzyColdDarkMatter}. An ALP field coupled to electromagnetism through a Chern-Simons term acts as a birefringent medium \cite{Carroll1990, Harari1990}. The polarization angle of the linearly polarized light propagating through the ALP field rotates by the amount proportional to the difference of the field values at emission and absorption. 
Since the ALP field oscillates at the Compton frequency, the polarization angle oscillates in time, and a broad experimental program searches for this signature: laboratory interferometers and optical cavities \cite{PhysRevD.98.035021, Obata2018, PhysRevD.100.023548, Nagano2019, PhysRevD.108.072005, PhysRevLett.132.191002, PhysRevLett.133.111003}, polarimetry of parsec-scale jets of active galactic nuclei \cite{Ivanov:2018byi, Huang2025}, protoplanetary disks \cite{Fujita2019, PhysRevD.110.103022, narita2026}, pulsars \cite{Liu2020, Liu2023, PhysRevD.111.062005, mptv-3x6g, vml6-f38z, yuwen2026meerkat}, supernova remnants \cite{ALP_taua_24, ALP_taua_26}, and the cosmic microwave background (CMB) \cite{Fedderke2019, ALP_BK12, ALP_BK14, ALP_SPT3g, ALP_PB}. These astronomical surveys, spanning from hours to decades, are sensitive to ultralight ALP \cite{Chigusa2020}.
It is worth noting that the observations of gravitational effects constrain the ALP mass and its fraction $\kappa$ of dark matter \cite{lyman_alpha_fuzzy, Nadler2021, PhysRevLett.126.071302, Chathirathas_2026}, whereas a measure of polarization angle oscillations constrains the combination $g_{a\gamma\gamma}\sqrt{\kappa}$, where $g_{a\gamma\gamma}$ is the ALP-photon coupling constant.  

In the standard halo model (SHM) \cite{PhysRevD.33.3495}, the ultralight galactic ALP field is a superposition of a number of plane waves with a Maxwellian velocity distribution and random phases. The resulting field is a stochastic Gaussian random field called a virialized ultralight field (VULF) \cite{Derevianko2018}. 
For astronomical polarization searches, an additional complication arises. The polarization oscillation signal depends on two widely separated points, the observer and the light source, and the light source is generally extended rather than a point.
This introduces two distinct effects: the spatial averaging of the field across the extended source, and the correlation between the observer and the source. Typically, the source is considered in one of two limits: either a point, or so large that the field average at the source vanishes. The intermediate regimes have not been described quantitatively. The observer-source correlation can make the signal amplitude sensitive to the source distance, so that distance uncertainties must be carefully taken into account.
These effects have not yet been organized within a common framework, and their treatment has varied across existing analyses. 

The focus of this paper is to provide a unified statistical framework for the polarization oscillation searches with arbitrary astronomical sources. We show that the statistics of the stochastic polarization oscillation amplitude can be classified into four regimes due to the two effects, and provide a seamless description including the transitional regimes between them. We also provide a likelihood that connects the observables to the ALP-photon coupling, and discuss implications for current and future observations.
Our analysis relies on a few assumptions. We assume an ALP with a single mass, and quantify the ALP signal statistics in the mass range from $10^{-23}$ to $10^{-18}~\mathrm{eV}$. 

The organization of this paper is as follows. Section~\ref{sec:signal_stats} presents the unified statistical formalism that describes the signal statistics for an astronomical extended sources. 
Section~\ref{sec:likelihood} describes the likelihood prescriptions and evaluates the confidence interval on the ALP-photon coupling constant.
Section~\ref{sec:searches} applies the formalism to the various experimental searches.
We conclude in Section~\ref{sec:conclusion}.

\section{Signal statistics} \label{sec:signal_stats}
We begin by reviewing the statistics of the ultralight dark matter field \cite{Derevianko2018, Centers:2019dyn}. We then apply this treatment to the signal statistics for the ALP-induced polarization oscillation of astronomical extended sources. We show that the source details are encapsulated in an effective amplitude, which is determined by two correlation factors, and we classify it into four regimes.

\subsection{Statistics of VULF}
During the formation of the Milky Way, the DM relaxes and virializes into a gravitational potential well and forms a virialized DM halo~\cite{PhysRevD.88.123517, Sikivie1992}, with a Maxwellian velocity distribution:
\begin{align}
    f_\mathrm{DM}(\vec{v}) = 
    \left( \frac{1}{2\pi \sigma_v^2} \right)^{3/2}
    \exp \left( -\frac{|\vec{v}-\vec{v}_\text{obs}|^2}{2\sigma_v^2} \right),
    \label{eq:SHM}
\end{align}
where $\sigma_v = 168~\text{km/s}\sim10^{-3}c$ is the per-axis velocity dispersion, and $\vec{v}_\text{obs} = 230~\text{km/s}$ is the velocity of the observer with respect to the galactic rest frame. 
In the SHM, the classical waves of unltralight DM are a superposition of plane waves with randomly distributed velocities and phases. Hereafter we refer to the resulting stochastic Gaussian random waves as a virialized ultralight field (VULF) \cite{Derevianko2018}.
The field oscillates at the Compton frequency
\begin{align}
    \omega_c \equiv \frac{m_a c^2}{\hbar} \simeq 4.8~\text{rad}~\text{year}^{-1}~\left(\frac{m_a}{10^{-22}~\text{eV}}\right),
    \label{eq:f_c}
\end{align}
and the reduced Compton wavelength sets the field's spatial scale
\begin{align}
    \lambda_c \equiv \frac{\hbar}{m_a c} \simeq 6.4\times10^{-3}~\text{pc}~\left(\frac{10^{-21}\,\text{eV}}{m_a} \right).
    \label{eq:lambda_c}
\end{align}
The velocity dispersion of the SHM leads to a spatial dispersion characterized by the coherence length 
\begin{align}
    \lambda_\text{coh} \equiv \frac{\hbar}{m_a \sigma_v} \simeq 1.2\times10^1~\text{pc}~\left(\frac{10^{-21}\,\text{eV}}{m_a} \right),
    \label{eq:lambda_coh}
\end{align}
and the spectral broadening characterized by the coherence time\footnote{Note that the coherence time and length are defined differently in the literature by a factor up to $2\pi$ \cite{Derevianko2018, Centers:2019dyn}.}
\begin{align}
    \tau_\text{coh} \equiv \frac{\hbar}{m_a \sigma_v^2} \simeq 7.8\times10^{4}~\text{year}~\left(\frac{10^{-21}\,\text{eV}}{m_a} \right).
\end{align}
The inverse of the coherence time corresponds to the line width of the ALP field $\sim \omega_c (\sigma_v / c)^2$.

A single plane wave comprising the VULF is
\begin{align}
    \phi_{\vec{v}}(\vec{r}, t) &= c_v\cos\left( \omega_v t -\hbar m_a^{-1} \vec{v}\cdot\vec{r}\right) \nonumber \\
    &\quad + s_v\sin\left( \omega_v t -\hbar m_a^{-1} \vec{v}\cdot\vec{r}\right),    
\end{align}
where $\omega_v = \omega_c(1 + v^2/2c^2)$ is the Doppler-shifted Compton frequency, $c_v$ and $s_v$ are Gaussian random variables with zero mean and variance 1/2. The VULF is a superposition of such random waves with the weight of Eq.~\eqref{eq:SHM}:
\begin{align}
    \phi(\vec{r}, t) 
    &= \frac{\hbar\sqrt{2\rho(\vec{r})}}{m_a c} \int d\vec{v} \sqrt{f_\mathrm{DM}\left(\vec{v}\right)} \phi_{\vec{v}}(\vec{r}, t),
    \label{eq:VULF_integral}
\end{align}
where $\rho(\vec{r})$ is the DM density.
The effective amplitude,
\begin{align}
    \Phi_0\equiv\frac{\hbar\sqrt{2\rho(\vec{r})}}{m_a c},
\end{align}
is determined by requiring the ensemble average of the time-averaged energy density to equal the DM density. Using $\Braket{c_v, c_{v^\prime}} = \Braket{s_v, s_{v^\prime}} = \delta(\vec{v} - \vec{v}^\prime)/2$ and $\Braket{c_v, s_{v^\prime}} = 0$, we obtain $\langle \phi(\vec{r}, t)^2 \rangle = \Phi_0^2/2$.
Since VULF is a superposition of random Gaussian variable, the VULF at each point is a random Gaussian variable with variance $\Phi_0^2/2$.
Using the SHM and the observation of the rotation curve in the Milky Way, a number of groups have estimated the local DM energy density to be $\rho$ = 0.3 - 0.55 GeV/cm$^3$ \cite{bovy2012local, McKee_2015, Sivertsson2018, PhysRevD.99.023012}.

In the mass scale considered here, the observation time is much shorter than $\tau_\text{coh}$, and the time evolution of the VULF reduces to a single-mode sinusoid as
\begin{align}
    \phi(\vec{r}, t) &\simeq \Phi_0 \left[ \phi_c \cos(\omega_c t) + \phi_s \sin(\omega_c t) \right]\\
    &= \Phi_0 \alpha \sin(\omega_c t + \theta)
    \label{eq:vulf_sine}
\end{align}
where $\phi_c$ and $\phi_s$ are Gaussian random variables with zero mean and variance 1/2. Equivalently, $\alpha$ is a normalized stochastic amplitude following a Rayleigh distribution with a scale factor of $1/\sqrt{2}$, for which $\langle \alpha^2 \rangle = 1$, and $\theta$ follows a uniform distribution between 0 and $2\pi$.

We now define the normalized two-point correlation function of the VULF. As we derive in Appendix~\ref{apx:tp}, for arbitrary two space-time points $(\vec{r}, t)$ and $(\vec{r}', t')$, the correlation function depends only on the separation as
\begin{align}
    \gamma(\vec{r} - \vec{r}', t-t') 
    &\equiv\frac{\Braket{ \phi(\vec{r}, t)\phi(\vec{r}', t') }}{\sqrt{\Braket{ \phi(\vec{r}, t)^2 }\Braket{ \phi(\vec{r}', t')^2 }}} \nonumber \\
    &=\frac{m_a^2c^2}{\hbar^2\sqrt{\rho(\vec{r})\rho(\vec{r}')}}\Braket{ \phi(\vec{r}, t)\phi(\vec{r}', t') }
    \label{eq:VULF_corr_definition}
\end{align}
For time separations $|\tau| \ll \tau_\text{coh}$, the normalized correlation function is 
\begin{align}
    \gamma(\vec{d}, \tau)
    = \exp\left(-\frac{|\vec{d}-\vec{v}_\text{obs}\tau|^2}{2\lambda_\mathrm{coh}^2}\right)
    \cos\left(\bar{\omega}\tau -\frac{m_a}{\hbar}\vec{v}_\text{obs}\cdot\vec{d}\right).
    \label{eq:VULFcorrelation}
\end{align}
where $\vec{d} \equiv \vec{r} - \vec{r}'$, $\tau \equiv t-t'$ and $\bar\omega \equiv \omega_c[1+(v_\text{obs}^2+3\sigma_v^2)/2c^2]$.
This two point function was first derived by \textcite{Derevianko2018}.
The coherence length controls the Gaussian envelope while the oscillatory factor is controlled by the Compton length $\lambda_c \sim 10^{-3} \lambda_\text{coh}$. This hierarchy is the origin of all the effects discussed below.

\subsection{Polarization oscillation of extended sources}
The polarization angle rotation induced by the ALP field depends only on the difference in the ALP field between the observer and the source:
\begin{align}
    \Delta\psi = \frac{g_{a\gamma\gamma}}{2}\Delta\phi,\quad\Delta\phi\equiv\phi_\mathrm{obs} - \phi_\mathrm{src},
\end{align}
where $g_{a\gamma\gamma}$ is the ALP-photon coupling constant \cite{Harari1990}.
In our mass range, the size of the observer is negligible compared with $\lambda_c$ and the observer can be treated as a point. 
On the other hand, the size of the light source can be non-negligible for astrophysical searches, and we need to consider the light emitted from a different point with a different retarded time. 
Without loss of generality, we set the observer at the origin, $\phi_\mathrm{obs} \equiv \phi(\vec{0}, 0)$, and the representative source position at $\vec d_\text{src}$.
With the normalized weight of the source region $w$,
the effective source field is
\begin{align}
    \phi_\mathrm{src} = \int d^3\vec{r}~ w(\vec{d}_\text{src} + \vec{r}) \phi(\vec{d}_\text{src}+\vec{r}, t_r),
\end{align}
where $\vec{d}_\text{src} + \vec{r}$ represents each point of the extended source and $t_r = -|\vec{d}_\text{src} + \vec{r}|/c$ is the retarded time.
Both $\phi_\text{obs}$ and $\phi_\text{src}$ are superpositions of Gaussian variables, so $\Delta\phi$ is also a Gaussian variable. 
As we derive in Appendix~\ref{apx:stochastic_amplitude}, its variance is
\begin{align}
    \Braket{(\Delta\phi)^2}
    &= \frac{\hbar^2}{m_a^2c^2} \left(\rho_\mathrm{obs} + \gamma_\mathrm{src}\rho_\mathrm{src} - 2\Gamma_{\mathrm{obs}, \mathrm{src}} \sqrt{\rho_\mathrm{obs} \rho_\mathrm{src}} \right),
    \label{eq:variance_diff}
\end{align}
where $\rho_\mathrm{obs}$ and $\rho_\mathrm{src}$ are the mean ALP densities at the observer and the source, respectively.
The $\gamma_\text{src}$ is the ``intra-correlation" within the light source and $\gamma_\text{obs, src}$ is the ``inter-correlation" between the observer and the light source
\begin{align}
    \gamma_\mathrm{src} &\equiv \iint d^3\vec{r}\,d^3\vec{r}'\,w\,w^\prime\gamma(\vec{r}-\vec{r}',\, t_r - t_r'), \label{eq:gamma_src} \\
    \Gamma_\mathrm{obs, src} &\equiv \int d^3 \vec{r}\,w\,
    \gamma(\vec{d}_\text{src} + \vec{r},\, t_r). \label{eq:gamma_obs_src}
\end{align}
We dropped the position dependence of $w$ for brevity.
As we derive later, the normalized inter-correlation is $\gamma_\text{obs, src}=\Gamma_\text{obs, src}/\sqrt{\gamma_\text{src}}$ then Eq.~\eqref{eq:variance_diff} can also be expressed as
\begin{align}
    \Braket{(\Delta\phi)^2}
    &= \frac{\hbar^2}{m_a^2c^2} \left(\rho_\mathrm{obs} + \gamma_\mathrm{src}\rho_\mathrm{src} - 2\gamma_{\mathrm{obs}, \mathrm{src}} \sqrt{\rho_\mathrm{obs} \gamma_\mathrm{src}\rho_\mathrm{src}} \right).
\end{align}
The measured polarization angle therefore oscillates as
\begin{align}
    \Delta\psi(t) = \frac{g_{a\gamma\gamma}}{2}\Phi\alpha \cos(\bar{\omega} t + \theta),~~\Phi^2/2 = \Braket{(\Delta\phi)^2}, \label{eq:dpsi}
\end{align}
where $\Phi$ is the effective signal amplitude, $\alpha$ is a normalized stochastic amplitude that follows the Rayleigh distribution with scale factor $1/\sqrt{2}$, and  $\theta$ is a random variable that follows uniform distribution from 0 to 2$\pi$. 
The effective signal amplitude $\Phi$ sets the characteristic signal strength and encapsulates the dependence on the source properties through the ALP density at the source and the two correlation factors $\gamma_\mathrm{src}$ and $\gamma_\mathrm{obs,src}$.

\subsection{Intra- and inter-correlation factors}
\begin{figure*}
    \centering
    \includegraphics[width = \textwidth]{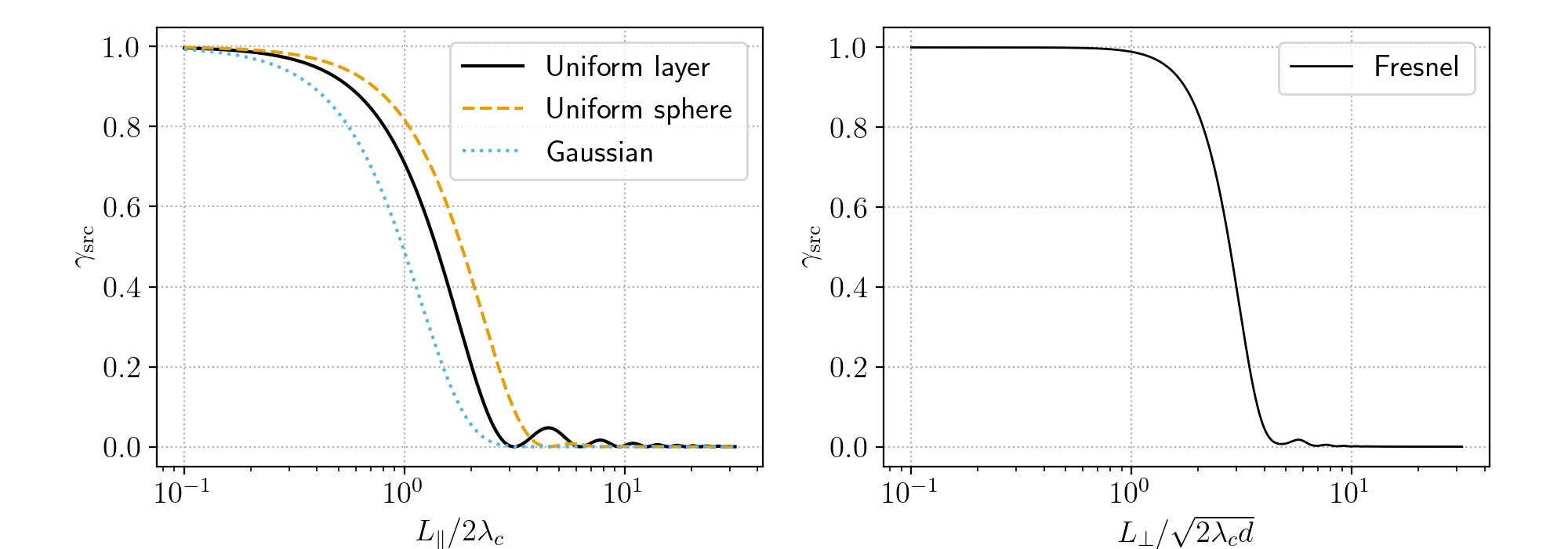}
    \caption{Left: Internal correlation $\gamma_\text{src}$ as a function of the line-of-sight depth of the source for various source profiles. 
    Right: Suppression of $\gamma_\text{src}$ due to the transverse extent of the source, described by a Fresnel integral. A uniform square transverse profile ($L_\perp \times L_\perp$) is assumed.
    }
    \label{fig:form_factor}
\end{figure*}
We evaluate $\gamma_\text{src}$ for a source with characteristic size of $L$ at distance $d \gg L$. We denote the line-of-sight depth of the source as $L_\parallel$ and transverse extent of the source as $L_\perp$. If the angular resolution of the measurement is smaller than the transverse extent of the source, we replace $L_\perp$ with the resolution, and the source can be regarded as an array of sources. 

We begin by considering the case $L \ll \lambda_\text{coh}$. The Gaussian envelope of Eq.~\eqref{eq:VULFcorrelation} is unity in this case, and Eq.~\eqref{eq:gamma_src} leads to
\begin{align}
    \gamma_\text{src} &= \iint d^3\vec r\,d^3\vec r'\,
    w\,w'\cos\!\left[\bar \omega \left( t_r - t_r' \right)\right].
    \label{eq:gamma_src_} 
\end{align}
We have dropped the spatial phase $\hbar^{-1} m_a \vec v_\mathrm{obs}\cdot( \vec r - \vec r')$, which is smaller than the retardation phase by $v_\mathrm{obs}/c \sim 10^{-3}$.
We expand the retardation phase as $c(t_r - t_r') \simeq r_\parallel - r_\parallel^\prime + (r_\perp^2 - r_\perp^{\prime 2})/2d$, where $r_\parallel$ is the line-of-sight offset and $r_\perp^2$ is the square of the transverse offset.
If the transverse extent satisfies the far-field limit:
\begin{align}
    L_\perp \ll  \sqrt{2\lambda_c d}, \label{eq:farfield}
\end{align}
the quadratic term of the retardation phase is negligible and Eq.~\eqref{eq:gamma_src_} simplifies to a one dimensional integral over the line-of-sight:
\begin{align}
    \gamma_\text{src} &\simeq \iint d r_\parallel\,d r_\parallel'\,
    w_\parallel\,w_\parallel'\cos\,\left[(r_\parallel - r_\parallel^\prime)/\lambda_c\right] \nonumber \\
    &\simeq \text{Re} \left[\int d r_\parallel~
    w_\parallel e^{i r_\parallel/\lambda_c}\int d r_\parallel~
    w_\parallel e^{-i r_\parallel/\lambda_c} \right] \nonumber \\
    &= \left| W \right|^2,
     \label{eq:gamma_src_ff}
\end{align}
where
\begin{align}
    W\equiv\int d r_\parallel~w_\parallel e^{i r_\parallel/\lambda_c}, ~~ w_\parallel \equiv \int d^2 \vec{r}_\perp w.
\end{align}
The intra-correlation is thus a squared Fourier transform of line-of-sight depth profile.
The left panel of Fig.~\ref{fig:form_factor} shows the internal correlations with simple profiles.
With $x \equiv L_\parallel/2\lambda_c$, a uniform layer with depth $L_\parallel$ gives $|W| \simeq \sin( x )/x$, a Gaussian profile of FWHM $L_\parallel$ gives $|W|\simeq\exp(-x^2 / 2\ln2)$, a uniform sphere profile gives $|W| \simeq 3j_1(x)/x$, where $j_1$ is the spherical Bessel function of the first kind.
The scale that controls the internal averaging is $\lambda_c$, and the dependence on the shape is small. 
A transversely large but very thin source retains $\gamma_\text{src}\simeq 1$ within the far-field limit. 
If the transverse profile or the angular resolution of measurement violates far-field limit, $w_\parallel$ turns into a Fresnel integral:
\begin{align}
    w_\parallel = \int d^2 \vec{r}_\perp~w e^{i r_\perp^2 /2d\lambda_c},
\end{align}
and suppresses $\gamma_\text{src}$ further as shown in the right panel of Fig.~\ref{fig:form_factor}.
Thus Eq.~\eqref{eq:gamma_src_ff} is an upper bound. 
Equation~\eqref{eq:gamma_src_ff} also applies to $L \gtrsim \lambda_\text{coh}$ as an upper bound\footnote{For the line-of-sight depth $L_\parallel$, the $\gamma_\text{src}$ always vanishes when $L_\parallel \gtrsim \lambda_c$, and Eq.~\eqref{eq:gamma_src_ff} can be extended to $L_\parallel \gtrsim \lambda_\text{coh}$. 
For the transverse extent $L_\perp$ there is an exception, although it is physically unlikely. 
For a very thin source with $L_\parallel  < \lambda_c \ll \lambda_\text{coh} < L_\perp < \sqrt{2\lambda_c d}$, the $\gamma_\text{src}$ will be suppressed due to the correlation of the VULF even in the far-field limit by approximately a factor of $L_\perp/\lambda_\text{coh}$ per axis.}.

The same expansion applies to the inter-correlation between observer and source.
Eq.~\eqref{eq:gamma_obs_src} leads to
\begin{align}
    \Gamma_\text{obs, src} 
    &\simeq \exp\left(-\frac{d^2}{2\lambda_\mathrm{coh}^2}\right) \int d^3\vec{r}\,w \cos\left( |\vec d_\text{src} + \vec r| / \lambda_c \right) \nonumber \\
    &\simeq \exp\left(-\frac{d^2}{2\lambda_\text{coh}^2}\right)
    \mathrm{Re}\left[e^{i d/\lambda_c}\,W(1/\lambda_c)\right] \nonumber \\
    &\simeq \exp\left(-\frac{d^2}{2\lambda_\text{coh}^2}\right) 
    |W|\cos\left(\frac{d}{\lambda_c} + \varphi_W \right),
     \label{eq:gamma_obs_src_ff}
\end{align}
where $\varphi_W = \arg(W)$, $|W|=\sqrt{\gamma_\text{src}}$ and the spatial phase is dropped again. 

We note that this formalism is also applicable to the extragalactic sources even though the SHM (Eq.~\eqref{eq:SHM}) depends on the parameters specific to the Milky Way. This is because $\gamma_\text{src}$ only assumes that the ALP field is a superposition of plane waves, and $\gamma_\text{obs, src}$ always vanishes, since $\lambda_\text{coh}$ must be smaller than the size of the Milky Way for the galaxy to form.


\subsection{Stochastic signal amplitudes in four regimes} \label{sec:stoch_amp}

\begin{figure}
    \centering
    \includegraphics[width = .49\textwidth]{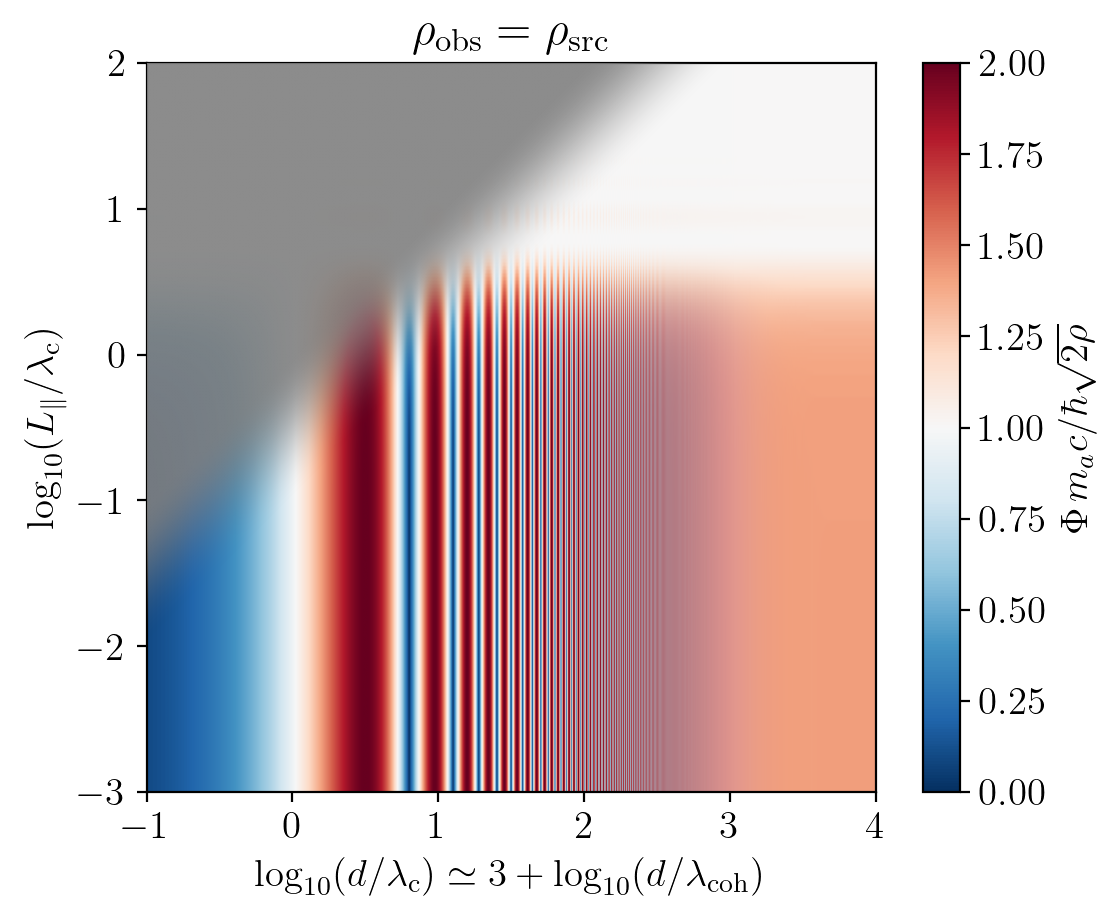}
    \caption{Normalized effective stochastic signal amplitude as a function of source distance $d$ and line-of-sight source depth $L_\parallel$. The same ALP density at the observer and the source, a uniform layer source profile and $\lambda_\text{coh} = 10^3 \lambda_c$ are assumed. The region where $L_\parallel \gtrsim d$ is excluded, as Eq.~\eqref{eq:variance_diff} does not apply there.
    The $(d /\lambda_c, L_\parallel / \lambda_c)$ plane can be split into four regimes: (1) the extended source regime $(L_\parallel/\lambda_c > 1)$; (2) the uncorrelated two-point regime $(L_\parallel/\lambda_c < 1$ and $d/\lambda_\text{coh} > 1)$; (3) the correlated two-point regime $(L_\parallel/\lambda_c < 1$, $d/\lambda_\text{coh} < 1$ and $1 < d/\lambda_c )$; (4) the small-scale regime $(L_\parallel/\lambda_c < 1$ and $d/\lambda_c < 1)$.
    The signal is dominated by the local observer field in regime~(1), while the observer and source contributions add in quadrature in regime~(2). Their interference produces alternating enhancement and suppression in regime~(3). Regime~(4) is the small-scale limit of this interference, in which the two contributions nearly cancel.
    The ALP mass $m_a$-axis runs diagonally in the $(d /\lambda_c, L_\parallel / \lambda_c)$ plane, since $m_a \propto \lambda_c^{-1}$. Depending on the source parameters $(d, L_\parallel)$, the $m_a$-axis runs through the four regimes differently and the $m_a$-axis splits into four regimes.
    }
    \label{fig:stoch_amp}
\end{figure}

We now classify the effective stochastic signal amplitude into four regimes. 
Fig.~\ref{fig:stoch_amp} shows the effective signal amplitude $\Phi$ (Eq.~\eqref{eq:dpsi}) in $(d, L_\parallel)$ space. This parameter space can be organized with three transition masses. The first one sets the boundary between the point-like source regime and extended source regime $(L_\parallel \sim \lambda_c)$, given by the line-of-sight depth of the sources $L_\parallel$:
\begin{align}
    m_a^L\simeq \frac{2\pi\hbar}{cL_\parallel}\simeq \SI{4.0e-23}{eV}\left( \frac{\SI{1}{pc}}{L_\parallel}\right),
\end{align}
below which the source can be regarded as a point. 

The second one determines whether the observer and the source are correlated $(d\sim\lambda_\text{coh})$:
\begin{align}
    m_a^d\simeq \frac{\hbar}{\sigma_v d}\simeq \SI{1.2e-20}{eV}\left( \frac{\SI{1}{pc}}{d}\right)
\end{align}
Since the correlation between the observer and the source turns on only when $\gamma_\text{src} \gtrsim 1$, the relation $m_a^d \leq m_a^{L}$ always holds in practice.

The last one marks the mass scale below which both the observer and the source lie within a single Compton wavelength of the ALP wave ($d \sim \lambda_c$):
\begin{align}
    m_a^c\simeq \frac{\hbar}{c d}\simeq \SI{6.4e-24}{eV}\left( \frac{\SI{1}{pc}}{d}\right)
\end{align}
With these three transition masses the ALP mass axis splits into four regimes: (1) the extended source regime $(m_a^L < m_a)$; (2) the uncorrelated two-point regime $(m_a^d < m_a < m_a^L)$; (3) the correlated two-point regime $(m_a^c < m_a < m_a^d)$; and (4) the small-scale regime $(m_a < m_a^c)$.

We now characterize the effective stochastic signal amplitude in each regime.
In the extended source regime (1), where $L_\parallel \gtrsim \lambda_c$ the internal correlation within the source vanishes ($\gamma_\text{src} = 0$) and the signal depends only on the local ALP field:
\begin{align}
    \Phi &\simeq \frac{\hbar}{m_a c}\sqrt{2\rho_\mathrm{obs}}.
    \label{eq:amp_extended_source}
\end{align}

In the uncorrelated two-point regime (2), where $d \gtrsim \lambda_\text{coh}$ the source can be treaded as a point source $(\gamma_\text{src}=1)$ and fields at the observer and the source are independent ($\gamma_\text{obs, src}$ = 0). Thus the two terms add in quadrature:
\begin{align}
    \Phi &\simeq \frac{\hbar}{m_a c}\sqrt{2(\rho_\mathrm{obs} + \rho_\mathrm{src})}.
    \label{eq:amp_uncorrelated_two_point}
\end{align}
The signal will be amplified using sources where the ALP density is high.

In the correlated two-point regime (3), where $\lambda_c < d < \lambda_\text{coh}$ the fields at the observer and the source are correlated $(\gamma_\text{obs, src} = \cos\varphi)$ and interfere:
\begin{align}
    \Phi &\simeq \frac{\hbar}{m_a c}\sqrt{2(\rho_\mathrm{obs} + \rho_\mathrm{src} - 2\cos \varphi \sqrt{\rho_\mathrm{obs} \rho_\mathrm{src}})},
    \label{eq:amp_correlated_two_point}
\end{align}
where $\varphi\equiv d/\lambda_c$.
The signal is largest when $d$ is an odd multiple of the $\pi\lambda_c$\footnote{In other words, the signal is largest when $m_a$ is an odd multiple of $\pi\hbar/dc$. This interference pattern is well-defined because the ALP line width $\sim m_a (\sigma_v/c)^2$ is much narrower than the spacing between adjacent maxima.}.
However, the astronomical distances $d$ are usually not known to the fraction of $\lambda_c$ (Eq.~\eqref{eq:lambda_c}). Therefore, the phase term $\varphi$ has to be treated as a nuisance parameter. We discuss its impact on parameter inference in Sec.\,\ref{sec:likelihood}.

Finally in the small-scale regime (4), where $d < \lambda_c$ the ALP density will be similar at the two points $(\rho_\text{obs} = \rho_\text{src})$ and we obtain
\begin{align}
    \Phi &\simeq \frac{2\hbar\sqrt{2\rho_\text{obs}}}{m_a c} \left\lvert\sin\left(\frac{d}{2\lambda_c}\right)\right\rvert.
    \label{eq:amp_small_scale}
\end{align}
The signal increases as the scale of the experiment, $d$, increases. This is a special case of the correlated two-point regime.

\section{Likelihood} \label{sec:likelihood}
\begin{figure*}
    \centering
    \begin{minipage}[h]{0.49\linewidth}
        \includegraphics[width = 0.99\textwidth]{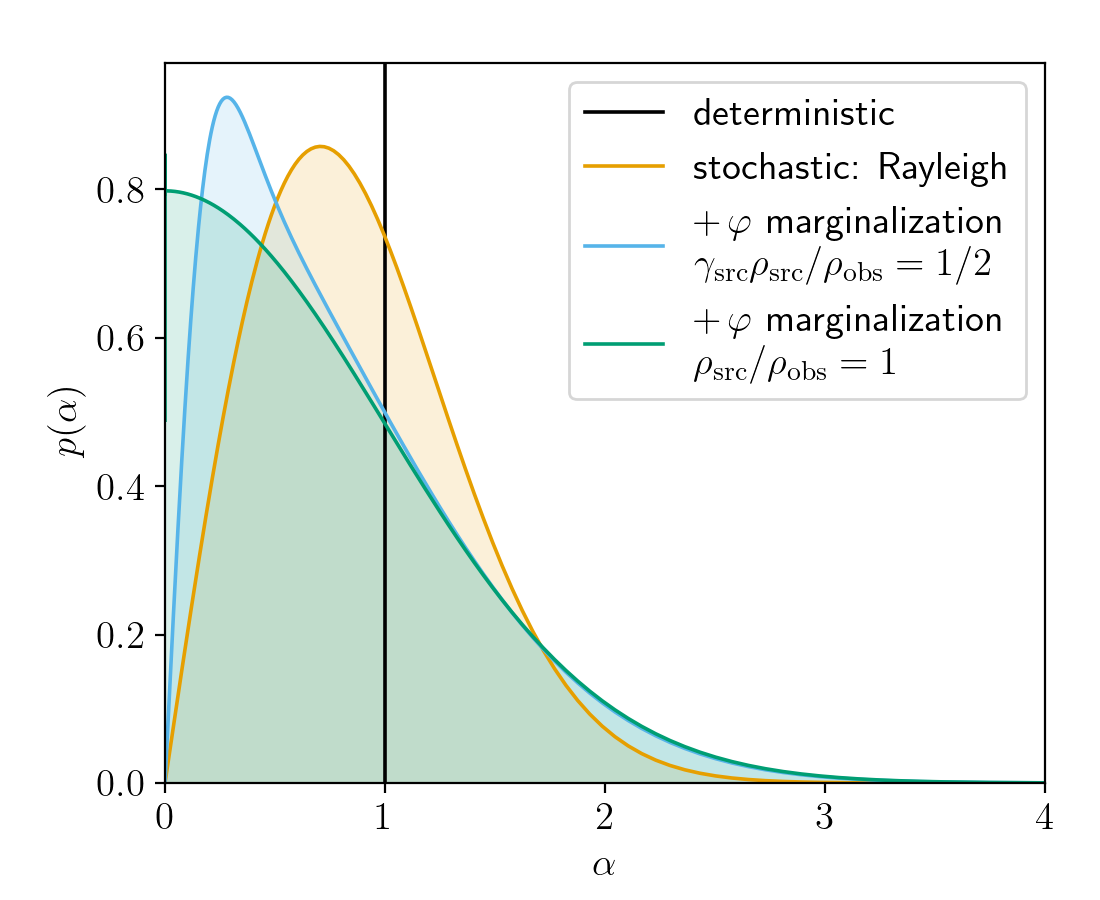}
    \end{minipage}
    \begin{minipage}[h]{0.49\linewidth}
        \includegraphics[width = 0.99\textwidth]{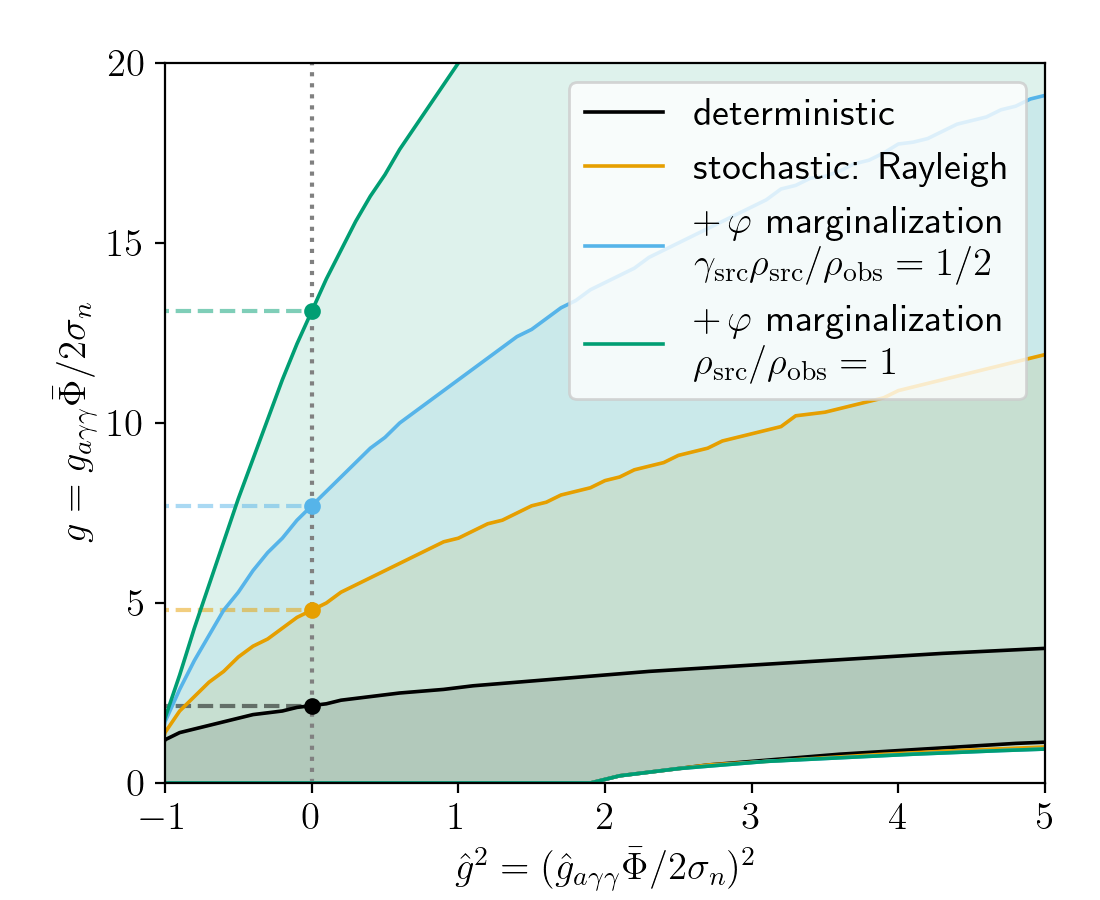}
    \end{minipage}
    \caption{
    Left: Comparison of the probability distribution of the normalized stochastic amplitude in the correlated two-point regime. 
    The black line shows the deterministic amplitude, which is a delta function. 
    The orange line shows the Rayleigh distributed stochastic amplitude of the Gaussian random field. The blue and green lines show the phase marginalized amplitude distribution with fully coherent case and half coherent case, respectively. The probability distribution reduces to a half-normal distribution in the fully coherent case. The heavier weight around zero leads to weaker upper bounds on the coupling constant $g_{a\gamma\gamma}$.
    Right: Comparison of Feldman-Cousins 95\% confidence interval of the normalized coupling constant $g$ as a function of $\hat{g}^2$.
    The horizontal dashed lines show 95\% upper limit of $g$ for observed value of $\hat{g}^2=0$.
    Colors are the same as in the left panel. 
    The marginalization over $\varphi$ weakens the 95\% upper limit on $g_{a\gamma\gamma}$ up to $\sim$ 3 with respect to the stochastic case. 
    }
    \label{fig:upperlimit}
\end{figure*}

Now we summarize how the stochasticity of the field propagates to the inference of the ALP-photon coupling constant $g_{a\gamma\gamma}$. 
We consider a time series of the polarization angles of the source reduced to the complex Fourier amplitude $d_f$ at frequency $f$, with Gaussian noise $n_f$\footnote{We define the complex amplitude such that a sinusoidal signal $\alpha \cos(2\pi f t + \theta)$ contributes to $\alpha e^{i \theta} = \phi_c + i \phi_s$ at the signal frequency $f$. }. We denote the variance of Gaussian noise as $\sigma_n^2 \equiv \langle |n_f|^2 \rangle $.
Since the VULF is also a Gaussian signal (Eq.~\eqref{eq:dpsi}), its contribution to $d_f$ is zero mean complex Gaussian with variance $g_{a\gamma\gamma}^2 \Phi^2 / 4$.
The stochastic signal enters the likelihood as an excess variance as
\begin{align}
    \mathcal{L}_\text{stoch}(d_f|\bm\theta; \varphi) = \frac{\exp(-|d_f|^2/\sigma_n^2(1 + g^2))}{\pi\sigma_n^2(1 + g^2)},
\end{align}
where 
\begin{align}
    g(\bm \theta; \varphi) \equiv \frac{g_{a\gamma\gamma}\Phi}{2\sigma_n} \label{eq:g}
\end{align}
is the normalized dimensionless coupling constant, and $\bm\theta = (g_{a\gamma\gamma}, m_a, \rho_\text{obs}, \rho_\text{src}, \gamma_\text{src})$ are the model parameters.
The maximum likelihood estimator of $\mathcal{L}_\text{stoch}$ is
\begin{align}
    \hat{g}^2 = \frac{|d_f|^2 - \sigma_n^2}{\sigma_n^2}.
    \label{eq:g2_hat}
\end{align}
This is unbiased because $\langle |d_f|^2 \rangle= \sigma_n^2 (1 + g^2)$.

For the astronomical two-point configuration, an additional complication arises in one of the regimes.
As discussed in Sec.~\ref{sec:stoch_amp}, in the correlated two-point regime $(m_a^c < m_a < m_a^d)$,
the phase $\varphi = d/\lambda_c$ in Eq.~\eqref{eq:amp_correlated_two_point} needs to be treated as a nuisance parameter, because the uncertainty of $d$ is typically much larger than $\lambda_c$. 
The probability distribution of the normalized stochastic signal amplitude is given by marginalizing over $\varphi$ contained in the scale factor of the Rayleigh distribution as:
\begin{align}
    \int \text{Rayleigh}\left(\alpha; \sigma=\frac{\Phi(m_a, \rho_\text{obs}, \rho_\text{src}, \gamma_\text{src}; \varphi)}{\sqrt{2} \bar\Phi}\right) p(\varphi) d\varphi,
    \label{eq:pdf_amp_marg}
\end{align}
where $p(\varphi)$ is a uniform prior for $\varphi$ and
\begin{align}
    \bar\Phi^2 = \int \Phi^2(m_a, \rho_\text{obs}, \rho_\text{src}, \gamma_\text{src}; \varphi) p(\varphi) d\varphi
\end{align}
is the phase averaged amplitude. In the correlated two-point regime, we replace $\Phi$ in Eq.~\eqref{eq:g} with the phase averaged amplitude $\bar\Phi$.

The left panel of Fig.~\ref{fig:upperlimit} compares the probability distribution of the normalized signal amplitude in the correlated two-point regime. Some works adopt a deterministic amplitude analysis, in which the ALP field amplitude is fixed to its expectation value. For the VULF as a Gaussian random field, however, the amplitude follows a Rayleigh distribution. As pointed out by \citet{Centers:2019dyn} the Rayleigh distribution has significant weight on small amplitudes, which weakens the constraining power on $g_{a\gamma\gamma}$.
After further marginalizing over $\varphi$, there is even larger weight on small amplitudes, which leads to even weaker constraints on $g_{a\gamma\gamma}$, as shown later.
This effect is most pronounced in the fully coherent case, $\gamma_\text{src}\rho_\text{src} / \rho_\text{obs} = 1$ and $\gamma_\text{obs, src} = \cos\varphi$. In this case, the scale factor reduces to $\sigma=|\sin(\varphi/2)|$, and the probability distribution (Eq.~\eqref{eq:pdf_amp_marg}) reduces to a half-normal distribution: $\sqrt{2/\pi} \exp(-\alpha^2/2)$.

We then consider the 95\% confidence interval on $g$ within the frequentist framework using the Feldman-Cousins approach \cite{Feldman_Cousins_1998}. 
The right panel of Fig.~\ref{fig:upperlimit} compares the 95\% confidence interval of $g$ as a function of $\hat{g}^2$.
The black and orange lines show the deterministic amplitude analysis and Rayleigh distributed stochastic analysis, respectively. These reproduce the result of \citet{Centers:2019dyn}, who found that the Rayleigh distributed stochastic analysis yields 95\% upper limits weaker than the deterministic analysis by a factor of $\sim$3 in a frequentist framework.
The additional marginalization over $\varphi$ further weakens the 95\% upper limits. 
When the observed value of $\hat{g}^2$ is zero, the 95\% upper limits in the fully coherent limit are factor $\sim 3$ weaker relative to the Rayleigh distributed stochastic analysis.
For a half coherent case, $\gamma_\text{src}\rho_\text{src} / \rho_\text{obs} = 1/2$ and $\gamma_\text{obs, src} = \cos\varphi$, the upper limits are factor $\sim1.5$ weaker relative to the Rayleigh distributed stochastic analysis.
The Bayesian approach reproduces similar results (Appendix~\ref{apx:bayesian}). 


\section{Implications for searches} \label{sec:searches}
\begin{figure*}
    \centering
    \begin{minipage}[h]{0.49\linewidth}
        \includegraphics[width = 0.99\textwidth]{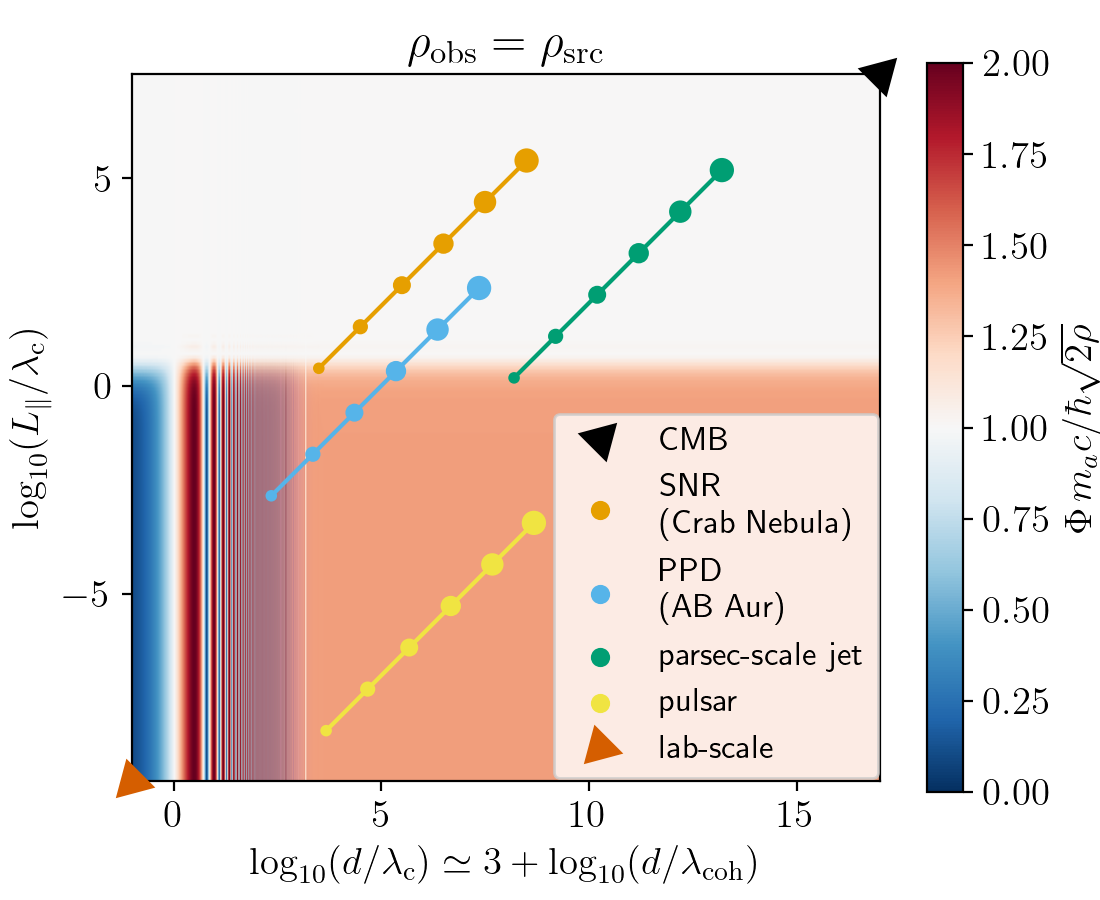}
    \end{minipage}
    \begin{minipage}[h]{0.49\linewidth}
        \includegraphics[width = 0.99\textwidth]{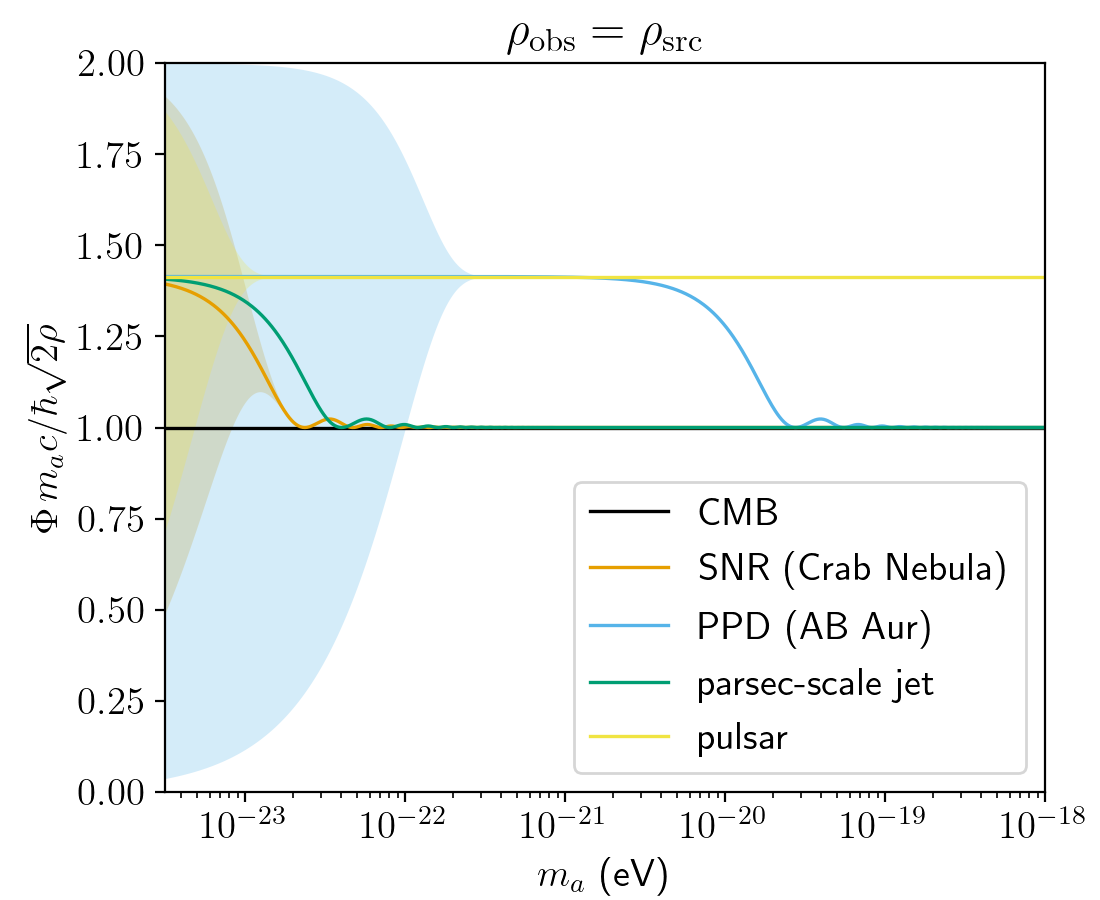}
    \end{minipage}
    \caption{
    Left: Normalized effective stochastic signal amplitude as a function of source distance $d$ and line-of-sight source depth $L_\parallel$. This figure extends the parameter range of Fig.~\ref{fig:stoch_amp} and additionally displays the $m_a$-axis for representative sources. Each line with markers represents the $m_a$-axis spanning from $10^{-23}$ to $10^{-18}$\,eV for each source, with markers placed at decade intervals. The CMB and lab-scale searches are out of the plotted range. 
    Right: Normalized effective stochastic signal amplitude for representative sources as a function of ALP mass $m_a$.
    This figure corresponds to a slice of the left panel along the $m_a$-axis of each source.
    Solid curves show $\sqrt{1+\gamma_\text{src}}$, which represents the amplification of the effective amplitude in point-like regime. The shaded region shows the envelope of the observer-source interference pattern determined by $\gamma_\text{obs, src}$, which is relevant in the correlated two-point regime. This interference pattern leads to a weaker constraint on $g_{a\gamma\gamma}$ when the source distance is uncertain.
    The CMB remains in the extended-source regime across the entire range. 
    The pulsars remain point-like across the entire range but transition to the correlated two-point regime.
    The parsec-scale jet transitions from extended-source regime to point-like source regime.
    The Crab Nebula and protoplanetary disk cross over two transition masses, $m_a^d$ and $m_a^L$.}
    \label{fig:searches}
\end{figure*}

We now evaluate the effective stochastic signal amplitude for representative sources.
Figure~\ref{fig:searches} compares the normalized stochastic signal amplitude in the typical mass range of astronomical searches from $10^{-23}$ to $10^{-18}$\,eV.
The corresponding time-scale is from hours to a decade.

The cosmic microwave background is the oldest observable light in the universe, emitted during the recombination era and polarized up to $\sim5\%$ due to Thomson scattering \cite{HU1997323}. The searches with CMB belong to the extended-source regime across the entire mass range, consistent with previous work \cite{Fedderke2019, ALP_BK12, ALP_BK14, ALP_SPT3g}.

Synchrotron radiation from electrons in magnetic fields is intrinsically polarized, and is emitted by a variety of astrophysical sources such as supernova remnants (SNR) and pulsars.
The Crab Nebula is one of the brightest supernova remnants. It has a diameter of $\sim\SI{1.7}{pc}$ \cite{Hester2008}, and $\sim\SI{2}{kpc}$ away \cite{Kaplan_2008}, leading to $m_a^L\simeq\SI{2e-23}{eV}$ and $m_a^{d} \sim \SI{6e-24}{eV}$.
This is consistent with the local field only signal model adopted in the previous work over the mass range from $10^{-22}$ to $10^{-18}\,\text{eV}$ \cite{ALP_taua_24, ALP_taua_26}, since this range lies within the extended-source regime. A decade of polarimetry will reach the point-like source regime, where the source term must be included in the signal amplitude.

The emission from pulsars originates within the light cylinder with size $L\lesssim 10^5\,\text{km}$ \cite{Philippov_2022}, and they can be regarded as point-like sources for $m_a\lesssim10^{-14}\,\text{eV}$. 
The pulsars are typically located at distances of a few kpc \cite{Manchester_2005}, leading to $m_a^d \sim \SI{5e-24}{eV}$. 
Thus the searches with pulsar polarimetry mostly remain in uncorrelated two-point regime, and Eq.~\eqref{eq:amp_uncorrelated_two_point} reproduces the signal model exploited with previous searches \cite{Liu2020, Liu2023, PhysRevD.111.062005, mptv-3x6g, vml6-f38z, yuwen2026meerkat}.
A correlated two-point regime will be applied for a decade of polarimetry or for closer pulsars.
Some of the current leading constraints on $g_{a\gamma\gamma}$ have been obtained by pulsar polarization arrays in the mass range we consider here. These limits are not expected to change significantly within our framework.

The protoplanetary disk (PPD) is often formed around a young star, and the light emitted from the young star and scattered in the PPD can be highly polarized \cite{Andrews_2020}. One of the precise measurements of PPD is performed for the AB Aur, which is $\sim144$\,pc away and the disk size is $\sim10^2-10^3$\,AU \cite{Hashimoto_2011}, leading to $m_a^L \sim \SI{3e-20}{eV} $ and $m_a^d \sim \SI{8e-23}{eV}$. The mass range spans over three regimes, and the degradation of constraining power on $g_{a\gamma\gamma}$ in correlated two-point regime applies to PPD. 
\citet{Fujita2019} and a subsequent studies \cite{PhysRevD.110.103022, narita2026} performed a deterministic amplitude analysis using the phase averaged amplitude for PPDs. Accounting for the transition masses and marginalizing over $\varphi$ will make upper limits more conservative.

The parsec-scale jets produced by active galactic nuclei are strongly linearly polarized extra-galactic sources \cite{Blandford2019}.  
For example, the 15~GHz VLBI observations of MOJAVE samples \cite{Lister_2021} are typically resolved in parsec-scale and their typical cosmological distances are $d\gtrsim10^8\,\text{pc}$. The observer-source correlation $\gamma_\text{obs, src}$ is negligible ($m_a^d \lesssim 10^{-28}\,\text{eV}$). For an emission region with a depth of $\sim\SI{1}{pc}$, the transition mass for the point-source regime is $m_a^L\sim\SI{4e-23}{eV}$. Therefore the signal amplification with source term (Eq.~\eqref{eq:amp_uncorrelated_two_point}) turns on for the measurements longer than several years.
\citet{Ivanov:2018byi} and subsequent study \cite{Huang2025} adopted the signal amplification due to $\rho_\text{src} = 20~\text{GeV/cm}^3~(\sim 50 \times \rho_\text{obs}$) across the wide mass range from $\sim10^{-23}\,\text{eV}$ up to $\sim10^{-20}\,\text{eV}$.
Depending on the details of the source profile, accounting for the source field suppression in the extended-source regime will make the upper limits more conservative.

The laboratory-scale broadband interferometric searches using polarization rotation effects operate in the small-scale regime \cite{PhysRevD.98.035021, Obata2018, PhysRevD.100.023548, Nagano2019, PhysRevD.108.072005, PhysRevLett.132.191002, PhysRevLett.133.111003}, and Eq.~\eqref{eq:amp_small_scale} reproduces the optical cavity response.

\section{Conclusion} \label{sec:conclusion}
We have presented a unified statistical description of ALP-induced polarization oscillations for light sources of arbitrary size and distance.
We showed that the source details are encapsulated in an effective signal amplitude, which is controlled by two correlation factors: the intra-correlation within the source $\gamma_\text{src}$ and the inter-correlation between the observer and the source $\gamma_\text{obs, src}$.
The relevant source scale is the line-of-sight depth in units of the Compton wavelength, and the distance in units of coherence length. The transverse extent is not relevant in the far-field limit.
We classified the effective signal amplitude into four regimes, and pointed out that the constraining power on $g_{a\gamma\gamma}$ degrades by factor up to three due to the distance uncertainty of the astronomical source, in the correlated two-point regime.
Our framework reproduces the ALP signal model of previous laboratory, pulsar, and CMB searches, and quantifies transitional regimes between them.

We applied our statistical framework to several representative astronomical sources and discussed its implications for current and future observations.
We find that the ALP field at the source typically becomes important for observations spanning approximately a decade, although pulsars are a notable exception. We also show that uncertainties in source distances must be carefully accounted for, particularly for nearby sources such as protoplanetary disks, since they can significantly affect the constraining power on $g_{a\gamma\gamma}$.
The presented framework provides a consistent basis for interpreting and comparing a broad range of ALP searches with astronomical polarized sources.

\section{Acknowledgements}
The author is grateful to Akito Kusaka and Lyman Page for providing valuable comments.
Claude Opus~5.5, a generative artificial intelligence tool developed by Anthropic, was used to proofread the manuscript. The author takes full responsibility for the content.

\appendix
\section{Derivation of the correlation function} \label{apx:tp}
We derive the normalized correlation function of the VULF Eq.~\eqref{eq:VULFcorrelation}.
Substituting Eq.~\eqref{eq:VULF_integral} into Eq.~\eqref{eq:VULF_corr_definition} and using $\Braket{c_v, c_{v^\prime}} = \Braket{s_v, s_{v^\prime}} = \delta(\vec{v} - \vec{v}^\prime)$ and $\Braket{c_v, s_{v^\prime}} = 0$, we obtain
\begin{align}
    \gamma(\vec{d}, \tau) = \text{Re}\int d^3\vec{v}~
    f(\vec{v}) \exp\left[i\left(\omega_{\vec{v}} \tau - \frac{m_a}{\hbar}\vec{v}\cdot\vec{d}\right)\right],
\end{align}
where $\omega_{\vec{v}} = \omega_c (1 + |\vec{v}|^2/2c^2)$ with $\vec{v} = \vec{v}_\text{obs} + \delta \vec{v}$.
The phase leads to
\begin{align}
    \omega_{\vec{v}}\tau - \frac{m_a}{\hbar}\vec{v}\cdot\vec{d} 
    &= \omega_c\tau\left(1 + \frac{v_\text{obs}^2}{2c^2}\right)
    - \frac{m_a}{\hbar} \vec{v}_\text{obs} \cdot \vec{d} \nonumber \\
    &\quad + \frac{m_a}{\hbar} \delta \vec{v} \cdot (\vec{v}_\text{obs}\tau - \vec{d}) + \omega_c \tau \frac{|\delta \vec{v}|^2}{2 c^2}.
\end{align}
Performing the Gaussian integral, we obtain
\begin{align}
    \gamma(\vec{d}, \tau)&=\text{Re}\Bigg\lbrace
    \exp\left[i\left(\omega_c(1+v_\text{obs}^2/2c^2)\tau-\frac{m_a}{\hbar}\vec{v}_\text{obs}\cdot\vec{d} \right)\right] \nonumber \\
    &\times\left(\frac{1}{1-i\tau/\tau_\text{coh}}\right)^{3/2}\exp\left( - \frac{|\vec{d} - \vec{v}_\text{obs}\tau|^2}{2\lambda_\text{coh}^2}\frac{1}{1-i\tau/\tau_\text{coh}} \right)\Bigg\rbrace.
\end{align}
The first order term of $(1-i\tau/\tau_\text{coh})^{-3/2}$ combines with the phase term into $\omega_c [1 + (v^2_\text{obs} + 3\sigma_v^2) / 2c^2]\tau$.
This reduces to Eq.~\eqref{eq:VULFcorrelation} when $|\tau| \ll \tau_\text{coh}$.

\section{Derivation of the effective amplitude} \label{apx:stochastic_amplitude}
We derive Eq.~\eqref{eq:variance_diff}. The variance of $\Delta \phi$ is given as
\begin{align}
    \Braket{(\Delta \phi)^2} = \Braket{\phi_\text{obs}^2} + \Braket{\phi_\text{src}^2} -2 \Braket{\phi_\text{obs}\phi_\text{src}}.
\end{align}
The variance of the local field is simply $\langle \phi_\text{obs}^2 \rangle = \hbar^2 \rho_\text{obs}/m_a^2c^2$.
The variance of the source-averaged field is
\begin{align}
    & \Braket{\phi_\text{src}^2} \nonumber \\
    &= \iint d^3\vec{r}\,d^3\vec{r}'\,w(\vec d_\text{src} + \vec r)w(\vec d_\text{src} + \vec r^\prime) \nonumber\\
    &\quad\times \Big\langle \phi(\vec d_\text{src} + \vec{r}, t_r)\phi(\vec d_\text{src} + \vec{r}', t_r') \Big\rangle \nonumber \\
    &= \iint d^3\vec{r}\,d^3\vec{r}'\,w\,w^\prime
    \nonumber\\
    &\quad\times
    \frac{\hbar^2\sqrt{\rho(\vec d_\text{src}+\vec{r})\rho(d_\text{src}+\vec{r}')}}{m_a^2 c^2}\,
    \gamma(\vec r - \vec{r}',\, t_r - t_r') \nonumber \\
    &\simeq\frac{\hbar^2}{m_a^2 c^2}\rho_\text{src} \gamma_\text{src}.
\end{align}
Here we assumed that the source is smaller than the scale of DM density variation, and approximated the ALP density around the source with the mean $\rho_\text{src}$.
The same calculation applies to the observer-source cross correlation: 
\begin{align}
    &\Braket{\phi_\text{obs}\phi_\text{src}} \nonumber \\
    &= \int d^3 \vec{r}\,w \Braket{\phi_\text{obs}\phi(\vec d_\text{src} + \vec{r}, t_r)} \nonumber \\ 
    &= \int d^3 \vec{r}\,w
    \frac{\hbar^2\sqrt{\rho_\text{obs}\rho(\vec d_\text{src} + \vec{r})}}{m_a^2 c^2}\,
    \gamma(\vec d_\text{src} + \vec{r}, t_r) \nonumber \\
    &\simeq \frac{\hbar^2 \sqrt{\rho_\text{obs}\rho_\text{src}}}{m_a^2 c^2} \Gamma_\text{obs, src}.
\end{align}
Thus, Eq.~\eqref{eq:variance_diff} holds.

\section{Bayesian upper limits on $g_{a\gamma\gamma}$} \label{apx:bayesian}
\begin{figure}
    \centering
    \includegraphics[width = .5\textwidth]{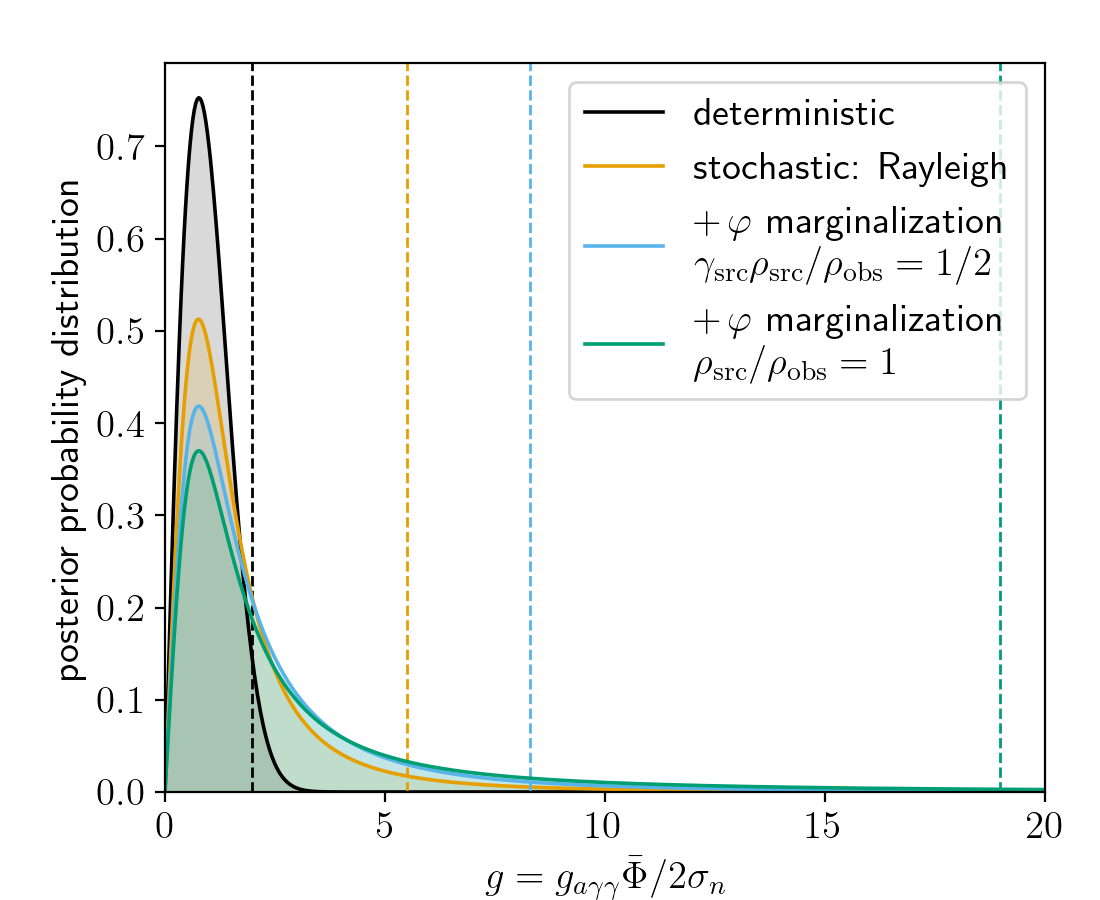}
    \caption{Comparison of posterior distributions and 95\% upper limits of $g$. The solid lines show posterior distribution with a subjective Jeffereys prior for $g$. The vertical dashed lines show 95\% upper limits. Marginalization over $\varphi$ weakens the 95\% upper limit up to $\sim$ 3 with respect to the stochastic case. }
    \label{fig:posterior}
\end{figure}

This appendix evaluates the Bayesian upper limits on $g_{a\gamma\gamma}$. This is a natural extension of the single-point treatment of \citet{Centers:2019dyn} to the astronomical two-point configuration.
Figure~\ref{fig:posterior} compares the posterior distribution and 95\% upper limits of normalized coupling constant $g$ with a subjective Jeffreys prior $(\propto g/(1+g^2))$ for $g$, following the method of \citet{Centers:2019dyn}.
\citet{Centers:2019dyn} evaluated the impact of signal stochasticity, and found that the 95\% upper limits of $g_{a\gamma\gamma}$ is factor 2-3 weaker with respect to the deterministic amplitude analysis.
When we additionally marginalize over $\varphi$, the Bayesian approach reproduces similar level of weakened upper limits as frequentist approach (Sec.~\ref{sec:likelihood}).

\bibliography{ref}

\end{document}